\documentclass[useAMS,usenatbib]{mn2e}
\usepackage{epsfig}
\usepackage{times}
  
\newif\ifAMStwofonts

\newcommand{\target}{Cen\,X--4}
\newcommand{\Msun}{$\rm \,M_{\odot}$}
\newcommand{\Rsun}{$\rm \,R_{\odot}$}
\newcommand{\Lsun}{$\rm \,L_{\odot}$}

\newcommand{\kms}{$\rm \,km\,s^{-1}$}
\newcommand{\kmsp}{$\rm \,km\,s^{-1}\,pixel^{-1}$}

\newcommand{\erg}{$\rm \,erg\,s^{-1}$}
\newcommand{\vsini}{$V_{\rm rot}\,\rm sin\,{\it i}$}
\newcommand{\ppm}{\,\pm\,}

\title[The spotty donor star in \target] 
{The spotty donor star in the X-ray transient \target}

\author[T.\,Shahbaz, C.A.\,Watson and V.S.\,Dhillon]
       {T.\,Shahbaz,$^{1,2}$  
\thanks{E-mail: tsh@iac.es}
\thanks{Based on observations collected at the European Southern 
Observatory, Chile, under the programme 081.D-0513(A)}
        C.\,A.\,Watson$^3$ and
        V.S.\,Dhillon$^4$ \\
$^1$Instituto de Astrof\'\i{}sica de Canarias (IAC), E-38200 La Laguna, Tenerife, Spain \\
$^2$Dept. Astrof\'\i{}sica Universidad de La Laguna (ULL), E-38206 La Laguna, Tenerife, Spain \\
$^3$Department of Physics and Astronomy, Queens University Belfast, Belfast BT7 1NN, UK \\
$^4$Department of Physics and Astronomy, University of Sheffield, Sheffield, S3 7RH, UK 
}

\pagerange{\pageref{firstpage}--\pageref{lastpage}}
\pubyear{2013}

\begin{document} 
\maketitle 
\begin{abstract} 

\noindent

We accurately determine the fundamental system parameters of the neutron--star 
X-ray  transient \target\ solely using phase-resolved high--resolution UVES 
spectroscopy. We first determine the radial-velocity curve of the 
secondary star and then model the shape of the phase-resolved absorption 
line profiles using an X-ray binary model. The model computes the exact 
rotationally broadened phase-resolved spectrum and does not depend on 
assumptions about the rotation profile, limb-darkening coefficients and 
the effects of contamination from an accretion disk. We determine the 
secondary star-to-neutron star binary mass ratio to be 
0.1755$\ppm$0.0025, which is an order of magnitude
more 
accurate than previous estimates. We also constrain 
the inclination angle to be 32$^\circ$$^{+8^\circ}_{-2^\circ}$. Combining these 
values with the results of the radial velocity study gives a neutron 
star mass of 1.94$^{+0.37}_{-0.85}$\Msun \ consistent with previous 
estimates. Finally, we perform the first Roche tomography reconstruction of the 
secondary star in an X-ray binary. The tomogram reveals surface 
inhomogeneities that are due to the presence of cool starspots. 
A large cool polar spot, similar to that seen in Doppler 
images of rapidly-rotating isolated stars
is present on the Northern hemisphere of 
the K7 secondary star and we estimate that $\sim$4 per cent of the total 
surface area of the donor star is covered with spots. This evidence for 
starspots supports the idea that magnetic braking plays an important 
role in the evolution of low-mass X-ray binaries.

\end{abstract}

\section{INTRODUCTION}
\label{intro}
 
In interacting binaries, such as dwarf novae and X-ray transients, the 
secondary star's rotational broadening (\vsini) combined with the radial 
velocity semi-amplitude ($K_{\rm 2}$) is normally used to determine 
the binary mass ratio of the system $q$ (=$M_2$/$M_1$: where $M_{\rm 1}$ and 
$M_{\rm 2}$ 
are the mass of the compact and secondary star, respectively). However, 
due to the faintness of the secondary star, its rotational broadening is 
usually determined by using intermediate resolution 
($\sim$0.5--1.0\,\AA) spectroscopy, where \vsini\ is typically 30--100\kms. With intermediate-resolution 
spectroscopy the information about the shape of the absorption lines is 
lost, and therefore the only information that can be extracted is the amount by which 
the stellar spectrum is broadened. The procedure commonly used to measure 
the secondary star's rotational broadening is to compare it to the 
spectrum of a slowly rotating template star, observed with the same 
instrumental configuration, that has been convolved with a limb-darkened 
standard rotation profile \citep{Gray92}. One assumes a limb-darkening 
coefficient for the spectral line and usually  adopts zero limb-darkening or the continuum value (which depends on the wavelength and 
the star's effective temperature). The width of the standard rotation 
profile is varied until an optimum match is found with the 
target spectrum \citep[see][]{Marsh94}.

\citet{Shahbaz98} showed that, in principle, it is possible to use the 
shape of the secondary star's absorption lines to determine the binary 
system parameters. In particular, one can determine the binary mass ratio 
directly by comparing the secondary star's line profile with a model 
line profile for the Roche-lobe filling secondary star geometry. Indeed,
\citet{Shahbaz03} presented a model \textsc{XrbSpectrum}, for determining 
the mass ratio of interacting binaries by directly fitting the observed 
line profile with synthetic spectra. The author makes direct use of 
\textsc{NextGen} model atmosphere intensities (Hauschildt, Allard, \& 
Baron 1999), which are the most comprehensive and detailed models 
available for cool stars. The model fully takes into account the varying 
temperature and gravity across the secondary star's photosphere, by 
incorporating the synthetic spectra into the secondary star's Roche 
geometry. As a result, \citet{Shahbaz98} determine the exact rotationally broadened 
line profile of the secondary star and so eliminate the need for a 
limb-darkening law, and the uncertainties associated with it.

X-ray transients  are  a subset of  low-mass X-ray  binaries (LMXBs) that 
display episodic, dramatic X-ray  and optical  outbursts,  usually lasting for
several months. Between outbursts, X-ray transients  remain in a quiescent
state, with typical X-ray luminosities less than $10^{32}$\erg, allowing the
optical detection of the faint low-mass donor star.  This allows the possibility
to perform radial velocity studies, probe the nature of the compact star and to
also determine its mass  \citep[e.g.][]{Charles06}.

The neutron star X-ray transient \target\ was discovered in 1969 by the 
Vela 5B satellite \citep{Conner69} when it went into X-ray outburst. 
During its second outburst in 1979 the optical counterpart was 
discovered \citep{Canizares80} and the mass accreting star was 
identified as a neutron star due to the fact that it displayed a Type I 
X-ray burst \citep{Matsuoka80}. After 1980 \target\ stayed in quiescence 
at $V\simeq 18.2$ and subsequent photometric studies led to the 
discovery of the orbital period $P_{\rm orb}$\,=\,15.1\,h 
\citep{Chevalier89} and the determination of the companion star's radial 
velocity curve \citep{Cowley88, MR90, Torres02, Avanzo05, Casares07}.  
Shahbaz, Naylor, \& Charles (1993) modelled the quiescent $H$-band light 
curve as due to the Roche-lobe filling secondary star's ellipsoidal 
modulation and obtained a binary inclination in the range 
$i$=31--54$^\circ$, assuming no contamination from the accretion disk. Using 
IR spectroscopy Khargharia, Froning, \& Robinson (2010) determined the 
contribution of the secondary star to the infrared flux and remodelled 
the $H$-band light curve of  \citet{Shahbaz93}, correcting for the 
fractional contribution of the donor star to obtain an inclination angle 
of 35$^\circ$$^{+4^\circ}_{-1^\circ}$

\citet{Torres02} obtained high-resolution optical spectroscopy of \target\ 
and  determined the mass function to be $f(M)$\,=\,0.220$\ppm$0.005\,\Msun, 
which was later refined by \citet{Avanzo05} to 
$f(M)$\,=\,0.201$\ppm$0.004\,\Msun.
\citet{Avanzo06} searched for the effects of irradiation on the 
absorption-line radial-velocity curve and concluded that there is no
evidence for irradiation.
\citet{Casares07} obtained much 
tighter constraints on $K_{\rm 2}$\,=\,144.6$\ppm$0.3\kms\ and $P_{\rm 
orb}$\,=\,0.6290522$\ppm$0.0000004\,d and also determined \vsini=44$\pm$3\,\kms
and $q$\,=\,0.20$\ppm$0.03, using the standard method by comparing the 
observed spectrum with a template star convolved with a limb-darkened 
rotation profile. Estimates for the secondary star's spectral 
type using optical spectroscopy include a K7\,V star \citep{Shahbaz93} and 
a K3--K5\,V star \citep{Torres02, Avanzo05}. \citet{Jonay05} determined 
$T_{\rm eff}$\,=\,4500\,K and  $\log\,g$\,=\,3.9, corresponding to a 
a K4\,V star, as well as the chemical abundances. 

The measurement of stellar masses in quiescent X-ray transients relies on
the determination of $K_{\rm 2}$ and $q$ from  spectroscopy    and $i$ normally
from photometry (via ellipsoidal modulation studies of the  optical/IR
lightcurves). This is mainly due to the fact that the secondary stars  in
quiescent X-ray transients are optically extremely faint, making it  difficult
for one to determine  accurate masses from spectroscopy alone. In non-eclipsing
systems the main uncertainty in the mass measurements is in the determination of
$i$.  If, in the future $i$ can be determined to a better accuracy, then one can
combine it with an accurate determination of $q$ to measure precise binary
masses.  In this paper we determine the  system parameters of  \target\
(primarily $q$ and $i$) by modelling the secondary star's Roche-lobe distorted 
absorption-line profiles with our X-ray binary model.  We also use for  the
first time the technique of Roche tomography to map the surface  inhomogeneities
on the secondary star in a LMXB. 

\begin{figure}
\hspace*{0mm}
\psfig{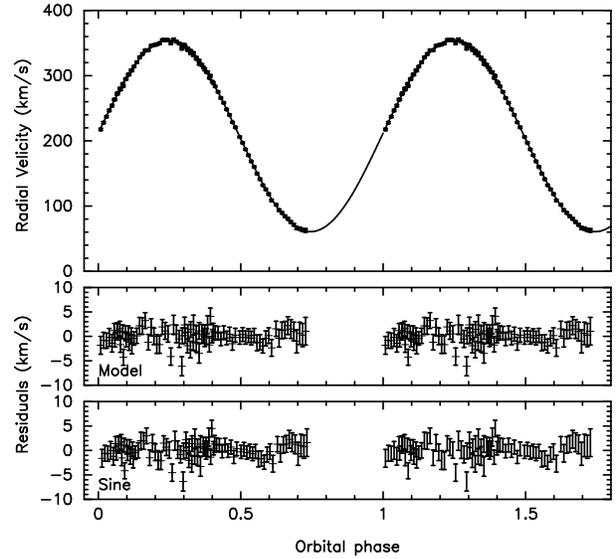}
\caption{The heliocentric radial-velocity curve of the secondary star in 
\target. The solid line shows a sinusoidal fit to the data. The middle a
 bottom panels show the residual after subtracting an  X-ray
 binary irradiation model and a
 circular orbit fit respectively. The data 
have been folded on the orbital ephemeris and are shown twice for 
clarity.}
\label{fig:rvcurve}
\end{figure}

\begin{table}
\caption{Log of VLT+UVES observations.}
\begin{center}
\begin{tabular}{lcc}\hline 
UT Date          &  UT range  & \# spectra \\
\hline
2008 June  08 & 01:05 -- 07:39 & 43 \\
2008 June  09 & 02:59 -- 04:38 & 12 \\
2008 June  13 & 00:16 -- 00:51 &  5 \\
2008 June  23 & 00:00 -- 05:06 & 35 \\
\hline       
\end{tabular}
\end{center}
\label{table:log}
\end{table}

\section{OBSERVATIONS AND DATA REDUCTION}
\label{obs}

We obtained spectra of \target\ in service mode during 2008 June with 
the UV-Visual Echelle Spectrograph (UVES) at the European Southern 
Observatory (ESO) Observatorio Cerro Paranal, using the 8.2 m Very Large 
Telescope (VLT). The UVES standard dichroic DIC1 was used yielding 
spectra covering the wavelength ranges 4727\,\AA\ to 5806\,\AA\ (hereafter green) 
and 5762\,\AA\ to 6837\,\AA\ (hereafter red).  In total 105 spectra of \target\ 
using an exposure time of 480\,s were taken as well as a spectrum of a 
K4\,V spectral-type template star (HD\,159341).  A log of the observations is shown in 
Table \,\ref{table:log}. A 0.8\arcsec\ slit was used resulting in a resolving 
power of 50,900 and an instrumental resolution of 5.9\kms\ 
measured from the full width at half-maximum (FWHM) of the arc lines. 
We used the UVES pipeline software which provides an absolute 
flux calibrated spectrum. The procedure consisted of bias subtraction, 
flat-fielding, wavelength calibration using thorium--argon lamps and absolute flux 
calibration. In what follows we only use 90 red spectra (the spectra 
taken on June 13 were not usable), because the signal-to-noise of the 
green spectra was poor due to the faintness of the source in this 
wavelength region.

\section{THE RADIAL VELOCITY CURVE AND MEAN SPECTRUM}
\label{spectrum}

Spectroscopic studies of the optical counterparts in quiescent   X-ray
transients include the determination of the  radial-velocity curve and the
rotational broadening  using the photospheric absorption lines arising  from the
companion star.  It has been known for some time, especially in studies of dwarf
novae and polars, that substantial heating of the secondary star shifts the
effective light centre of the secondary away from the centre of mass of the
star \citep{Davey92}.
 This results in a significant distortion of the radial-velocity curve, 
leading to a biased semi-amplitude and a non-circular radial-velocity
curve \citep{Shahbaz00}. \citet{Avanzo06} studied in detail the possible
effects of irradiation on the radial-velocity curve of \target\ and concluded
that there are no such effects,  not surprisingly given  the very low  quiescent
X-ray (and UV) luminosity  ($L_X<5\times 10^{32}$\erg; \citealt{Cackett13}). 

To determine the radial-velocity curve of \target\  we normalized the individual
\target\ spectra and template star  spectra by dividing through by a first-order
polynomial fit and then  subtracting a high-order spline fit to carefully
selected continuum  regions. This ensures that the line strength is preserved
along the  spectrum, which is particularly important when the absorption lines
are  veiled by differing amounts over a wide wavelength range.  We corrected 
for radial velocity shifts by cross correlating the individual \target\  spectra
with the template K4\,V star (rotationally broadened by 44\kms\  to match the 
rotational velocity of the secondary star),  using the method of 
\citealt{Tonry79}) and regions  devoid of emission and interstellar lines.  The
radial-velocity curve  i.e. the radial-velocity after correcting for the
systemic  radial-velocity of the  template star which was found to be 12.8\kms,
measured from the position of the H$\alpha$  absorption line) is shown in
Fig.\,\ref{fig:rvcurve}.

A circular  orbit fit gives the following parameters; 
$\gamma$\,=\,194.5$\ppm$0.2\kms, $K_{\rm 2}$\,=\,147.3$\ppm$0.3\kms,  $P_{\rm
orb}$\,=\,0.629059$\ppm$0.000017\,d and  $T_{\rm
0}$\,=\,HJD\,2454626.6214$\ppm$0.0002, where $T_{\rm 0}$ is time at  phase 0.0
defined as inferior conjunction of the secondary star and  $\gamma$ is the
systemic velocity (1--$\sigma$ errors are quoted with the  error bars rescaled
so that the reduced $\chi^2$ of the fit is 1.).  We also fit the radial
velocity curve with our X-ray binary model (see section\,\ref{model}), which
includes the effects of irradiation. The  $\chi^2$ for the circular orbit
fit and the X-ray binary model fit  are 87.2 and 81.6 with 87 and 86 degrees of
freedom, respectively;  the residuals of the circular orbit and irradiation
model fit are shown in  shown in Fig.\,\ref{fig:rvcurve}.  An F-test concludes
that the circular orbit model and the irradiation model are indistinguishable. 
Therefore, like   \citet{Avanzo06} we also conclude
that there are no effects of irradiation on the absorption lines in \target.

\begin{figure}
\psfig{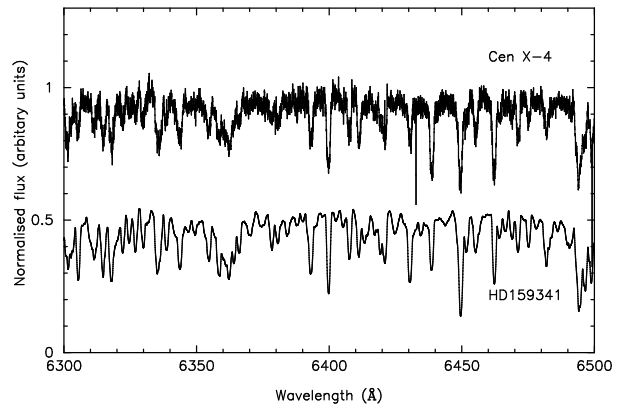}
\caption{The Doppler averaged spectrum of \target\ (top) and the K4\,V 
template star HD\,159341 (bottom) rotationally broadened by 44\kms.}
\label{fig:red_spectrum}
\end{figure}

The 
velocity--corrected spectra (using the circular orbit fit) were then combined to produce a 
variance--weighted Doppler averaged spectrum. The Doppler 
averaged spectrum of \target\ and the template star spectra were then 
binned onto the same uniform velocity scale (1.6\kmsp). The final spectrum 
of \target\ has a signal-to-noise ratio of $\sim$45 per pixel in the continuum and 
is shown in Fig.\,\ref{fig:red_spectrum}.

\section{THE BINARY MASS RATIO AND INCLINATION ANGLE}
\label{qi}

To determine the fundamental binary parameters, one would 
ideally like to observe a time-series of high signal-to-noise unblended 
absorption lines at high spectral resolution. The Least-Squares 
Deconvolution (LSD) method \citep{Donati97a} allows us to do exactly 
this, since it effectively stacks the thousands of stellar absorption 
lines in an echelle spectrum to produce a single 'average' absorption-line 
profile with increased signal-to-noise ratio; theoretically the 
increase in signal-to-noise ratio is the square root of the number of 
lines observed.  It has been used in the spectropolarimetric 
observations of active stars \citep[e.g.][]{Donati97b} and in Doppler 
imaging studies \citep[e.g.][]{Barnes04}. LSD has also been used in 
conjunction with Roche tomography to map the surface brightness 
distribution of the secondary stars in cataclysmic variables (CVs; see 
\citealt{Watson07} and references within) and to 
compute high-quality line profiles of X-ray binaries \citep{Shahbaz03, 
Shahbaz07}.

LSD assumes that all of the absorption lines are rotationally broadened 
by the same amount, and hence just requires the position and strength of 
the observed lines in the echelle spectrum to be known. We generated a 
line list appropriate for a K4\,V star ($T_{\rm eff}$\,=\,4500\,K, 
$\log\,g$\,=\,4.0) from the Vienna Atomic Line database 
(\citealt{Kupka00}). Approximately 5500 lines were used in the 
deconvolution process. Our version of LSD propagates the errors through 
the deconvolution process \citep[for details see][]{Watson07}.

LSD profiles were constructed for each of the individual 
Doppler-corrected \target\ spectra using regions devoid of emission 
lines and interstellar lines. We also computed the LSD profile for the 
K4\,V template star (HD\,159341). In the next sections, $q$  
is determined using two different methods. The first method uses 
the standard determination of \vsini\ that, when combined with $K_{\rm 
2}$, gives $q$. The second method involves the determination of $q$ 
directly using a Roche-lobe filling star model.

\begin{figure}
\psfig{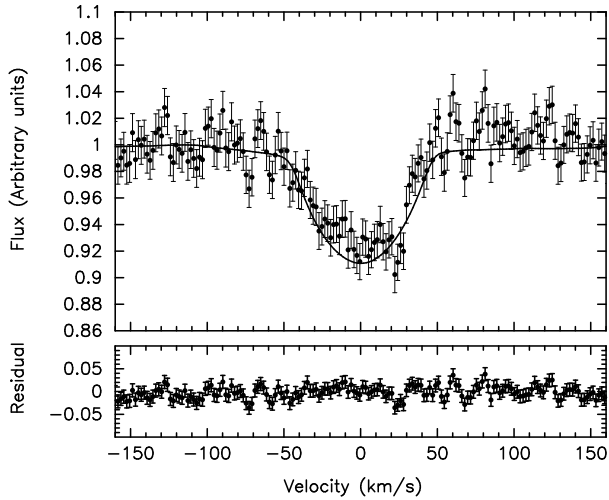}
\caption{The determination of \vsini\ for a LSD profile of \target\ at 
orbital phase 0.0 using the standard method (see section \ref{vsini}), 
where one compares the 
observed spectrum with a template star convolved with a limb-darkened 
Gray rotation profile. The solid line shows the LSD profile of 
the K4\,V template star rotationally broadened by 44.5\kms\ using a 
limb-darkened spherical rotation profile scaled to match the 
\target\ LSD profile. The bottom panel shows the residual of the fit 
after the optimal subtraction.}
\label{fig:vsini_lsd}
\end{figure}

\subsection{Spherical rotation profile}
\label{vsini}

The secondary stars in X-ray binaries are tidally locked and in 
synchronous rotation, therefore for a given orbital period the width of 
the line profile scales with the size of the star's Roche-lobe. It is 
easy to show that \vsini\ and $R_{\rm 2}$ (the radius of a sphere whose 
volume is the same as the volume of the secondary star) are related 
through the expression

\begin{equation}
\it V_{\rm rot}\,{\rm sin}\,i/K_{\rm 2}\,=\,(1+q)\,R_{\rm 2}/a, 
\label{eqn1}
\end{equation}

\noindent
where $a$ is the binary separation \citep{Horne86}. Thus by measuring 
\vsini\ and $K_{\rm 2}$ one can 
determine $q$. Previous models for the spectra of Roche-lobe filling 
secondary stars assume that the rotational broadening of their spectra 
can be modelled by convolving the spectrum of a non-rotating star with a 
line-broadening function for a slowly rotating spherical star, e.g. the Gray 
function. However, the use of the Gray profile to describe the spectra 
of Roche-lobe filling stars produces significantly biased measurements 
of the rotational velocities in interacting binaries. Convolution also 
gives biased rotational velocities even for rapidly rotating single 
stars \citep{Collins95}. Using equation\,\ref{eqn1} to convert from 
rotational velocity to mass ratio leads to a 5 per cent systematic 
underestimate in the mass ratio when using the relations from 
\citet{Paczynski71} or \citet{Eggleton83} for $R_{\rm 2}/a$, because 
they assume a spherical secondary star (Welsh, Horne, \& Gomer 1995, 
\citealt{Marsh94}). The unknown limb darkening coefficient in the line 
when applying the Gray broadening profile to measure \vsini\ also 
introduces a systematic bias as large as 14 per cent. This is because of the 
uncertainty about the non-spherical secondary stars radius
and the
fact  that the projected radius and hence rotational broadening and $q$ are 
strongly dependent on orbital phase $\phi$ \citep{Welsh95}.  Thus the  correct
expression for \vsini\ should be

\begin{equation}
\rm V_{\rm rot}(\phi)\,sin\,i/K_{\rm 2}\,=\,(1+q)\,R_{\rm 2}(\phi)/a 
\label{eqn2}
\end{equation}

\noindent
and, as pointed by \citet{Welsh95}, the spherical approximation can still 
be used provided that one models the phase-resolved spectra and not the 
mean spectrum.

\begin{figure}
\psfig{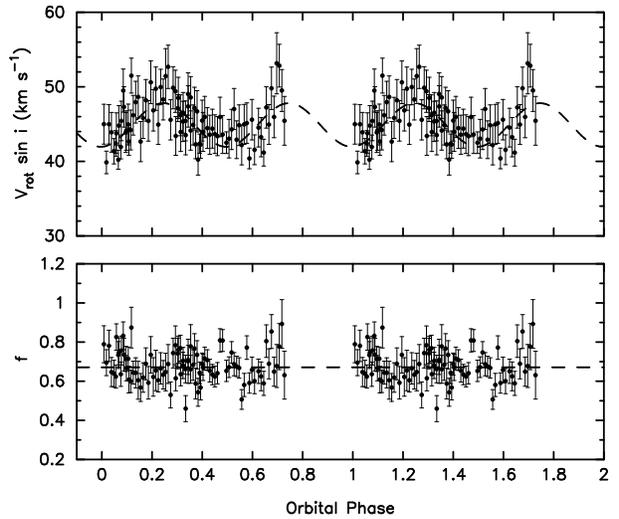}
\caption{The results of the phase-resolved measurements of \vsini\ (top 
panel) and $f$ (bottom panel) in \target, determined using the LSD 
profiles of \target\ and the standard method, where one compares the 
observed spectrum with a template star convolved with a limb-darkened 
standard Gray rotation profile. The data are shown twice for 
clarity.}
\label{fig:vsini_var}
\end{figure}

To determine \vsini\ the phase-resolved LSD profiles of \target\ were 
compared to a rotationally-broadened version of the template star LSD 
profile using the standard optimal subtraction procedure 
\citep{Marsh94}.  We optimally subtracted a constant (representing the fraction of 
light from the template star $f$) multiplied by a rotationally broadened 
version of the template star, using a spherical rotation profile 
\citep{Gray92} with a linear limb-darkening coefficient ($u$\,=\,0.72; 
\citealt{Claret00}) appropriate for a K4\,V star at the central 
wavelength of the red spectra. The optimal subtraction was performed 
over the line profile from -100 to 100\kms. The $\chi^2$ of the fit was calculated 
and the optimal rotational broadening and factor $f$ were determined for 
each orbital phase.  The LSD spectrum of \target\ at phase 0.0 and 
the rotational-broadening template-star fit is shown in 
Fig.\,\ref{fig:vsini_lsd}; 
the $\chi^2$ of the fit is 136.7 (132 degrees of freedom).
The values determined for \vsini\ and $f$ are 
shown in Fig.\,\ref{fig:vsini_var}, where the 1-$\sigma$ errors have been 
rescaled so that the reduced $\chi^2$ of the fit is 1.

A Roche-lobe filling star should show variations in \vsini\ with orbital 
phase as the secondary star rotates, in a similar manner to the 
ellipsoidal variations (due to the star's changing projected 
area), which show a double-humped modulation on half the orbital period. 
A fit to the phase-resolved rotational velocities with a sinusoidal 
modulation with half the orbital period gives a mean rotational velocity 
of \vsini\,=\,44.9\kms\ with a semi-amplitude of 2.9$\ppm$0.3\kms. It 
should be noted that the analysis above assumes that the limb darkening 
coefficient appropriate for the radiation in the lines is the same as 
for the continuum.  However, in reality this is not the case, and the 
absorption lines in late-type stars will have core limb-darkening 
coefficients much smaller than that appropriate for the continuum 
\citep{Collins95}. Therefore we also perform the analysis using zero 
limb-darkening and obtain a similar rotational velocity modulation but 
with a mean \vsini\,=\,41.0\kms\ and a semi-amplitude of 
2.9$\ppm$0.3\kms. Using the peak-to-peak values obtained using continuum 
and zero limb darkening we obtain a \vsini\ range of 38.1 (=\,41.0-2.9) 
to 47.8 (=\,44.9+2.9) \kms, which when combined with $K_{\rm 2}$ in 
equation 1 gives $q$ in the range 0.139--0.234. Note that our 
determination of $q$ here takes into account the phase-dependent values of 
\vsini\ and the uncertain line limb-darkening. \citet{Casares07} and 
\citet{Torres02} obtained $q$ values which are consistent with ours 
using the mean spectrum, 
though it should be noted that they did not take into account
account the phase dependent values of \vsini.

The fractional contribution of the secondary star should mimic the 
ellipsoidal variations. We do not, however, find any obvious sinusoidal 
modulation in the fractional contribution. This may be due to short-term 
flaring events dominating changes in the veiling factor of the 
absorption lines -- it is well known that such flaring events are 
present in all quiescent X-ray transients \citep{Zurita03} as well as in 
\target\ \citep{Chevalier89, Shahbaz10}. These flares are present at 
different levels during the different quiescent states (see 
section\,\ref{dis:lcurve}).

\begin{figure}
\psfig{angle=0,width=8.0cm,file=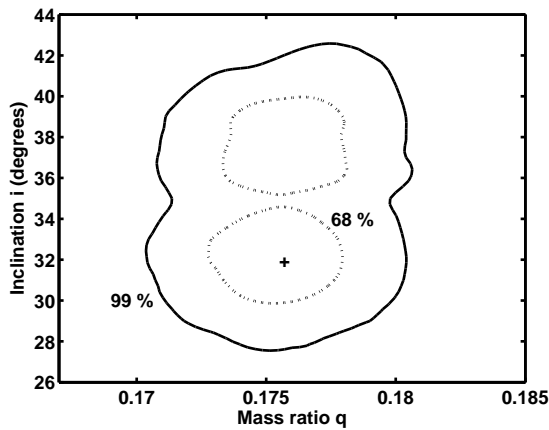}
\caption{The result of the model LSD profile fits to the phase 
resolved LSD profiles of \target\ in the ($q$-$i$) plane. The best fit 
at $q$\,=\,0.1755 and $i$\,=\,32$^\circ$ is shown with a cross and the 68 and 99 per 
cent confidence regions are show as the dotted and solid lines, respectively. }
\label{fig:qi}
\end{figure}

\subsection{Roche-lobe rotation profile}
\label{model}

In order to obtain an accurate determination of $q$ that does not depend 
on assumptions about the rotation profile and limb-darkening 
coefficients, we use the model \textsc{XrbSpectrum} described in 
\citet{Shahbaz03}. The model computes the exact rotationally-broadened 
model spectrum, which can then be directly compared with the observed 
spectrum of \target.  The parameters that determine the Roche geometry 
of the secondary star in an interacting binary are $q$, $i$, $K_{\rm 
2}$, $T_{\rm eff}$ the Roche-lobe filling factor and the gravity 
darkening coefficient. We assume that the secondary fills its Roche 
lobe, a safe assumption in X-ray binaries given that we observed an 
accretion disk \citep{Chevalier89,Shahbaz96}.  The width of the lines are primarily determined by $q$ 
and $K_{\rm 2}$, whereas $i$ and $\beta$ (the gravity darkening 
coefficient) mainly determine the shape of the absorption line. The 
gravity darkening coefficient was assumed to be 0.08, which is 
appropriate for 
late-type stars with a convective envelope \citep{Lucy67, Sarna89}.  
The velocity, gravity and temperature for each element varies 
across the star due to the shape of the Roche-lobe. For each visible 
element the model computes the local effective temperature, gravity 
and limb angle and then determines the specific intensity using the 
\textsc{NextGen} model atmosphere intensities \citep{Hauschildt99}. 
For a specified wavelength (limited to the range 6300--6800\,\AA) 
the model then integrates the visible specific intensity 
values over the visible surface  of the Roche-lobe and 
gives the exact rotationally-broadened 
spectrum of the secondary star at a given orbital phase. The intrinsic 
profile has a FWHM of 9.5\kms, which is set by the velocity scale of the 
\textsc{NextGen} model atmospheres (4.7\kmsp). The model is similar to 
the program \textsc{LinBrod} which uses \textsc{atlas9} and 
\textsc{moog} for its atmosphere and spectrum synthesis \citep{Bitner06}.

In order to compare the model spectrum with the observed spectrum, we 
computed the model spectrum at the same orbital phases and 
taking into account all sources that can velocity broaden the spectra. 
The spectra of \target\ consists of 90 Doppler corrected spectra 
covering orbital phases 0.00 to 0.73 with exposure times of 480\,s. We 
first computed the model spectrum for a given value of $q$, $i$, $K_{\rm 
2}$, $T_{\rm eff}$ and orbital phase.  We then allowed for the smearing 
of the absorption lines due to the motion of the secondary star during 
the length of each exposure at each orbital phase, by convolving the 
model spectrum with a rectangular function. We also convolved the 
spectrum with a Gaussian function with FWHM equal to the instrumental 
resolution of the data. We also convolved the observed LSD profiles with 
the intrinsic line profile shape of the \textsc{XrbSpectrum} model 
spectra so that we can compare the model and observed LSD profiles, 
taking into account all sources of velocity broadening. 
Finally, given the poorer velocity dispersion of the model 
spectra (4.7\kmsp) compared to the observed spectra (1.6\kmsp), all the 
data and models were binned onto the same uniform velocity scale of 
4.7\kmsp.

The model spectra were then normalized using a continuum spline fit and  then
LSD profiles were determined in the wavelength range  6300\,\AA\--6500\,\AA\
(note that we are limited in wavelength by the  model synthetic spectra). We
then compared each phase-resolved  model LSD line profile to each observed LSD
line profile by optimally  subtracting a scaled version of the line profile from
the observed line  profile (to allow for any variable disc contribution) and
computing the $\chi^2$ of the fit. This optimal subtraction  was performed over
the line profile from -100 to 100 \kms. We then sum the  $\chi^2$ values obtained for
all orbital phases.

\begin{figure*}
\psfig{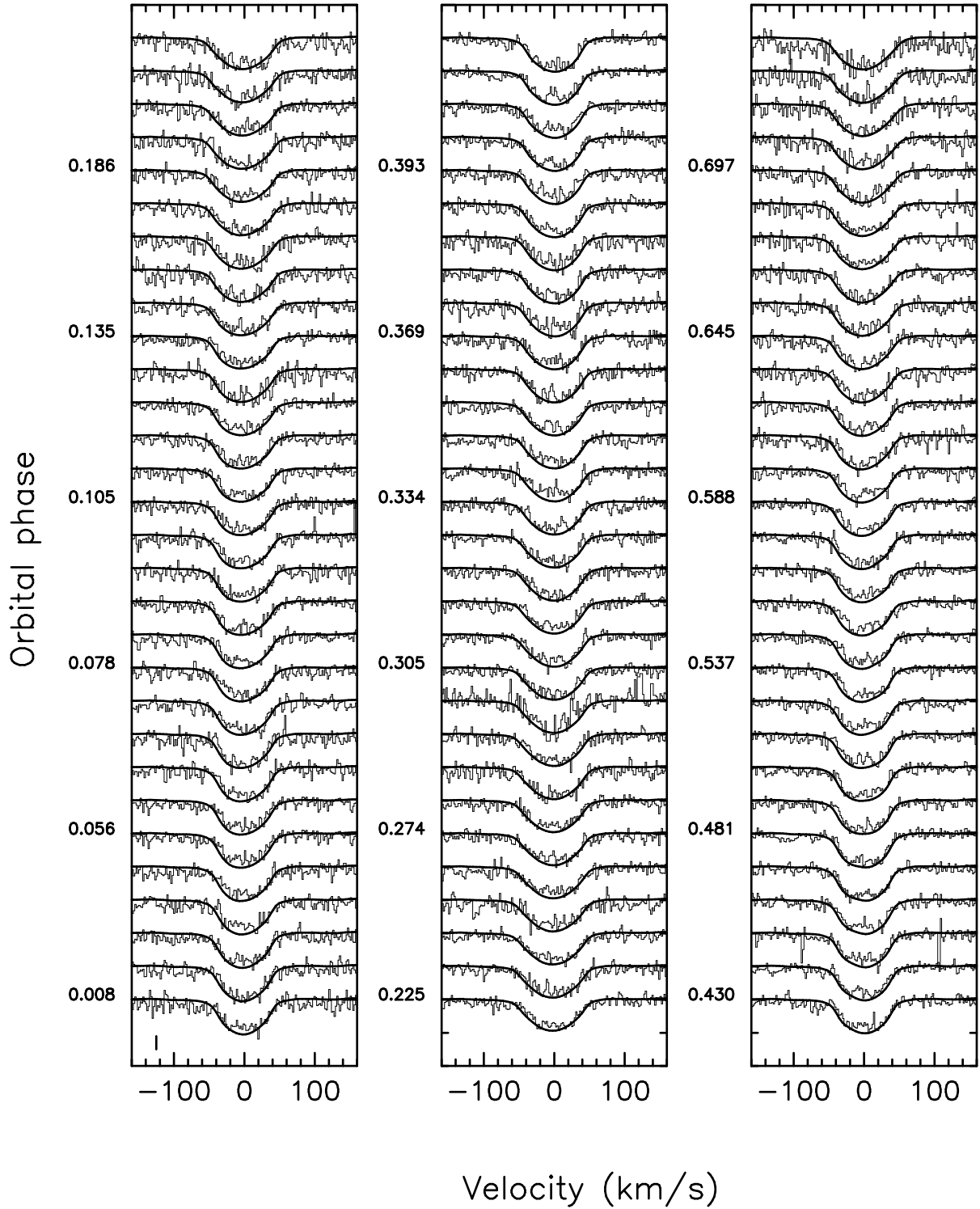}
\caption{The best-fit model LSD profiles (thick line) with 
the observed LSD profiles of \target\ (thin lines). 
The orbital motion has been removed assuming the binary parameters found in
Section\,\ref{spectrum} and the data have been shifted  vertically by 0.1 
(continuum units) for 
clarity. The typical uncertainty ($\pm 1 \sigma$) in the data 
is shown in the bottom 
left corner.}
\label{fig:data_model}
\end{figure*}

Computing the models is computationally expensive (taking $\sim$ 2 hrs 
of CPU time for a given $q$, $i$ and orbital phase) and so we perform a 
preliminary grid search in the ($q$-$i$) plane; $q$ in the range 0.16 to 
0.20 in steps of 0.01 and $i$ in the range 30$^\circ$--50$^\circ$ in steps of 
2.5$^\circ$. A gravity darkening coefficient of 0.08 was used 
throughout, the same as that has been used by previous authors 
to fit the optical/infrared light curves 
\citep{Khargharia10} and $T_{\rm eff}$ and $K_{\rm 2}$ were fixed at 
4500\,K \citep{Jonay05} and 147.3\kms\ (see 
Section\,\ref{spectrum}), respectively. Once a minimum $\chi^2$ was 
found we increased the resolution of the grid to 0.002 and 1$^\circ$ in 
$q$ and $i$, respectively. 
The best fit 
resulted in a minimum $\chi^2$ of 3800 (3780 degrees of freedom)
at $q$\,=\,0.1755 and 
$i$\,=\,32$^\circ$; 68 and 99 per cent confidence levels are 
shown in Fig,\,\ref{fig:qi} (the contours have been rescaled so that the 
reduced $\chi^2$ of the fit is 1). As one can see $q$ is well 
constrained to within 0.0025, whereas $i$ is only constrained to 
$^{+8^\circ}_{-2^\circ}$ (68 per cent uncertainties). 
In Fig,\,\ref{fig:data_model} we show the individual LSD profiles of 
\target\ with the best fit model profiles. As one can see the model fits 
the data reasonably well at most orbital phases.

\section{THE BINARY MASSES}
\label{masses}

Using our determined values for  $P_{\rm orb}$, $K_2$, $q$ and $i$ we can determine the binary masses using  the mass function  
$P_{\rm orb} K_{\rm 2}^3 / 2 \pi G\,=\,M_{\rm 1} \sin^3 i / (1+q)^2$.
In order to determine the uncertainties in $M_{\rm 1}$ and $M_{\rm 2}$ we  used a
Monte Carlo simulation, in which random values for the  observed quantities
were drawn that follow a Gaussian distribution, with mean and  variance the same as the observed values.   For parameters with symmetric  uncertainties $P_{\rm orb}$, $K_{\rm 2}$ (see Section\,\ref{spectrum})  we assume the distribution to be Gaussian.
We use  the ($q,i$) solutions found in Section\,\ref{model} to determine the actual
distribution for $q$ numerically by calculating the maximum likelihood distribution 
using the actual $\chi^2$ values, for a given value of $i$.  We then determine the 
cumulative  probability distribution of the asymmetric distribution numerically and 
then pick random values for the probability to obtain random values for the parameter. We obtain 
$M_{\rm 1}$\,=\,1.94$^{+0.37}_{-0.85}$\,\Msun\ and 
$M_{\rm 2}$\,=\,0.34$^{+0.07}_{-0.15}$\,\Msun, consistent with previous measurements
\citep{Shahbaz93,Torres02,Casares07,Khargharia10}.
Given the measured masses and orbital period, we use  Kepler's Third Law to  
determine $a$. Eggleton's  expression for the effective radius of the 
Roche-lobe \citep{Eggleton83}  then determines $R_{\rm 2}$, which with 
$T_{\rm eff}$\,=\,4500\,K inferred from the spectral type \citep{Gray92} 
and Stefan-Boltzmann's law gives the star's luminosity $L_{\rm 2}$.
We obtain $R_{\rm 2}$\,=\,0.99$^{+0.06}_{-0.18}$\Rsun, 
$L_{\rm 2}$\,=\,0.36$^{+0.04}_{-0.12}$\Lsun\ and 
$a$\,=\,4.07$^{+0.24}_{-0.72}$\Rsun.
 
\begin{figure*}
\psfig{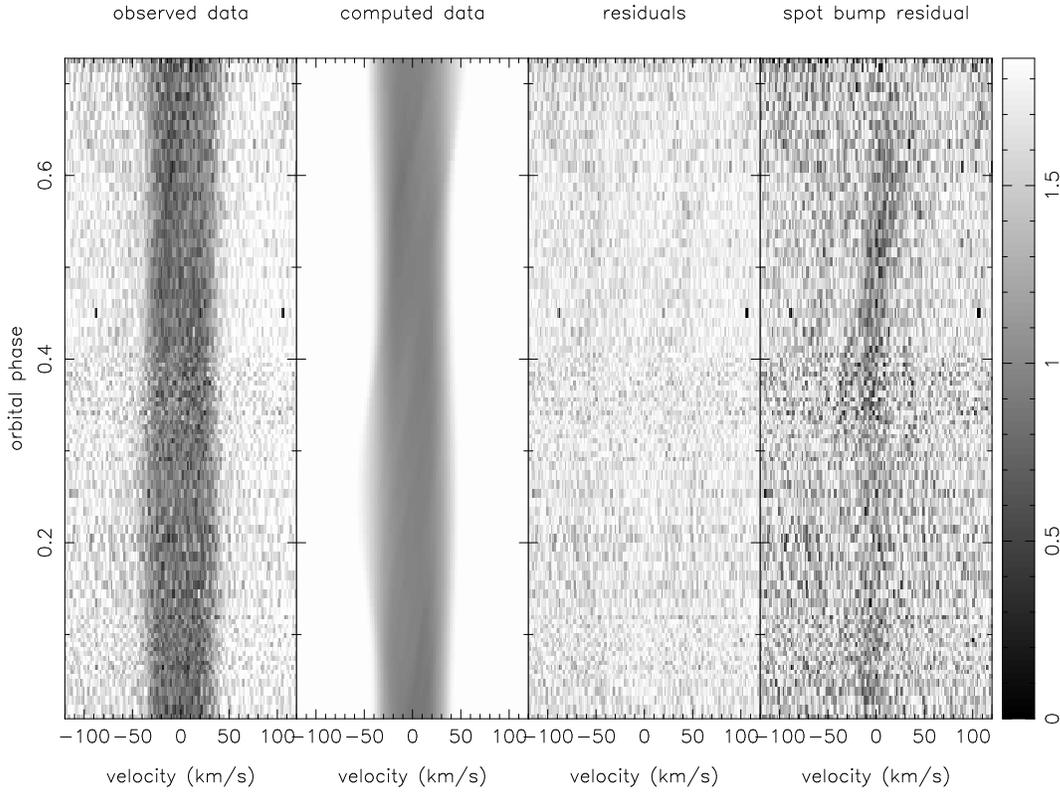}
\caption{
Trailed spectra of \target\ showing (from left to right), the observed LSD data,
computed data from the Roche tomography reconstruction, and the residuals.
Starspot features in these three panels appear bright. A greyscale intensity
wedge is also shown, where a value of  1 corresponds to the maximum line depth
in the reconstructed profiles. In the final (right-hand) panel we plot the
residuals after a theoretical line-profile assuming an 
unspotted  stellar surface was
subtracted from each LSD profile in order to enhance starspot features. In this
case, spot features now appear dark in the plots, and the contrast has been
stretched by a factor of 1.5 to further enhance the spot features. The orbital
motion has been removed assuming the binary parameters found in
Sections\,\ref{spectrum} and \ref{model}, which allows the individual starspot
tracks across the profiles and the variation in \vsini\ to be more clearly
observed. 
}
\label{fig:roche_trail}
\end{figure*}

\section{EVIDENCE FOR STARSPOTS}
\label{spots}

It is well known that non-uniform intensity distributions on the donor 
star surface can introduce systematic errors in the determined binary 
parameters due to the fact that the centre-of-mass and the 
centre-of-light of the donor may no longer be coincident. So we explored 
the dataset for the signatures of stellar activity, such as starspots.

Roche tomography is an astro-tomographic technique that images the 
surface of the secondary star in interacting binaries \citep{Rutten94, 
Watson02}. The procedure is fully tested and artefacts caused by 
systematic errors are well characterized and understood. For a detailed 
review of the technique see \citet{WD01, DW01, WD04}. In summary, Roche 
tomography uses phase-resolved spectral-line profiles to produce a map 
of the line-intensity distribution across the secondary star's surface 
in three-dimensional real space.  It is analogous to Doppler imaging of 
single stars, but with the difference that the star is assumed to be 
Roche-lobe filling, rotates around the binary centre-of-mass, is tidally 
locked, and has undergone orbital circularization.

While Doppler Tomography \citep{MH88} is a frequently used tool for mapping the
components in CVs and X-ray binaries in velocity space, Roche tomography has the
advantage that it can deal with effects such as self-obscuration and
limb-darkening of the donor star, and is better suited to studies of the stellar
components in such systems.  When performing Roche tomography it is important to
use the correct system  parameters ($i$, $q$, $M_{\rm 1}$ and $M_{\rm 2}$), as
if incorrect system parameters are adopted then these will introduce spurious
artefacts in the final tomogram \citep{WD01}. The system parameters are normally
obtained by constructing {\em entropy landscapes}, where reconstructions to the
same $\chi^2$ are carried out for different combinations of component masses,
and the entropy obtained in the final map for each pairing is plotted on a grid
of $M_{\rm 1}$ versus $M_{\rm 2}$. Since entropy encodes the information 
content in the map, and adopting incorrect system parameters introduces spurious
artefacts in the reconstruction, the parameters that yield the map with the
least information content and therefore least structure (maximum entropy) due to
artefacts are selected.

We constructed a series of entropy landscapes (see \citet{Watson07} and
references therein for a detailed description of the technique) for various
orbital inclinations in order to provide an independent estimate of the system
parameters. Since Roche tomography treats limb-darkening in a different manner
to the  \textsc{XrbSpectrum} model analysis outlined in Section\,\ref{model},
this was carried out for the case of no limb-darkening and also assuming a
square root limb-darkening law with coefficients of $c$ = 0.522 and $d$ = 0.362
(consistent with the tabulated values from \citet{Claret98} for a star with
$T_{\rm eff}$ = 4500\,K and $\log g$ = 4.0 centred on the $R$-band). Our initial
results gave mass ratios of $q$ = 0.197$ \pm$ 0.011 in the case of no
limb-darkening, and significantly higher mass ratios when limb-darkening was
assumed -- inconsistent with that found in the \textsc{XrbSpectrum} analysis.
($q$=0.1755$\ppm$0.0025). 
This discrepancy was tracked down to a combination of the relatively low
signal-to-noise of the dataset for Roche tomography purposes, coupled with a
systematic noise bias in the continuum introduced by the technique. In detail,
the maximum-entropy algorithm that is implemented in Roche tomography requires
that all data are positive. For absorption lines, this means that the profile is
first inverted and then all remaining negative data-points are set to zero. This
results in a positive ``biasing'' of the noise in the continuum regions, and for
low signal-to-noise data an improved fit can be found by artificially inflating
the Roche-lobe so that the positive noise near the line-wings can be fit as well
-- leading to a systematically higher value for $q$. In order to surmount this
problem, we have added a ``virtual pixel''  to the reconstruction process that
contributes a constant value to all velocity bins at all phases. This injects an
offset to the data that is then fit within the image reconstruction process,
with the result that initially negative data points no longer need to be
cropped. The addition of a virtual pixel has previously been implemented for low
signal-to-noise data in Roche tomography by \citet{Watson03}, where we found a
similar impact on the determination of the component masses.

After implementing this correction, we found that while we were unable to
reliably constrain the inclination, the mass ratios found at $i$=32$^{\circ}$
bracketed those determined by the \textsc{XrbSpectrum} analysis 
($q=$0.1755$\ppm$0.0025). In the case of
no limb-darkening, we find $q$=0.168, whereas for the root limb-darkening case
we find $q$=0.195. We suspect that the better treatment of limb-darkening by 
\textsc{XrbSpectrum} leads to a better binary parameter constraint in this case.
For low signal-to-noise data, Roche tomography may additionally be biased as it
attempts to fit features due to noise -- something that  \textsc{XrbSpectrum} is
less prone to.

\begin{figure}
\hspace{5mm}
\psfig{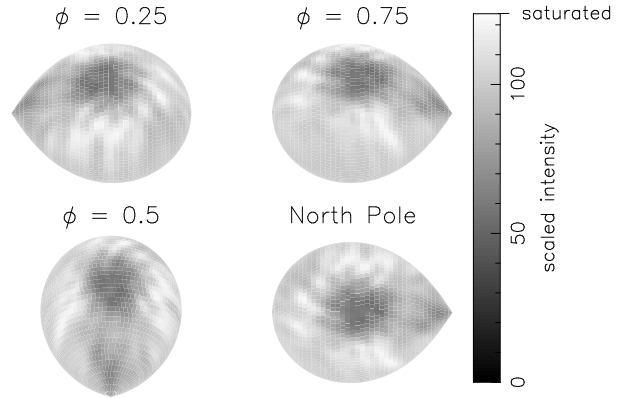}
\caption{
Roche tomogram of \target. The intensity scale is such that dark scales
represent regions of lower absorption and as such both starspots and any 
regions of ionising irradiation would both appear dark.
The map is shown as the observer would see it at an inclination 
of $i$\,=\,32$^\circ$ except for the lower right 
image, where the observer is directly above the north pole.
}
\label{fig:roche_map}
\end{figure}

In Fig.\,\ref{fig:roche_trail} we show the results of the  Roche tomography
reconstruction of the secondary star in \target\ using the LSD profiles. 
The data and corresponding fit are displayed as well as the residuals. 
To allow the reader to see the spot features more clearly, we also show
the residuals after a theoretical line-profile assuming a blank 
unspotted  stellar surface is 
subtracted from each LSD profile in order to enhance starspot features. 
The Roche tomogram is shown in
Fig.\,\ref{fig:roche_map}. The most notable feature in the Roche tomogram is a
large, dark starspot feature which lies in the  Northern polar region; 
the dark region at the inner Lagrangian point is due
to gravity darkening, as expected for a Roche-lobe filling star
\citep{Shahbaz98}.
This can
also be seen in the trailed spectra as a feature  that runs through the spectra
remaining near the core of the lines, and  is visible throughout the orbital
cycle (as expected for a high-latitude  feature seen on a star with a relatively
low inclination angle). 

While the Roche tomogram shows a number of features, such as small-scale  dark
regions and streaks at all latitudes, the majority of these are due  to the
projection of noise in the dataset. Indeed, the spectra are only  just of
sufficiently high signal-to-noise to confidently recover large  features that
have high visibility (such as the polar spot feature). We  are, however,
confident in the detection of the polar spot as it is both  visible in the
trailed spectra, and also covers a far larger surface  area than can be
attributed to noise (which creates artefacts on a much  smaller scale).
Calculating robust parameters for the polar spot, such  as its filling-factor
and  area, is made difficult due to the low  signal-to-noise of the dataset
relative to that obtained in other spot  maps of CV donors reconstructed with
Roche Tomography \citep{Watson03,Watson06,Watson07}.
Therefore, we  have only attempted to calculate an approximate
spot coverage of the  polar feature -- which we find covers $\sim$4 per cent of
the total  surface area of the donor star. The true polar-spot coverage is
likely  to be higher, since only the most  prominent (darkest) spots can be
confidently detected in this dataset,  and regions of lower spot-filling factors
will be difficult to discern.  Nonetheless, this is an appreciable size, and
indicates that LMXB donor stars,   like in CVs, also have high global
spot coverages \citep{Watson07}. This Roche tomogram therefore  represents the first concrete
observational evidence of strong magnetic activity  on the donor stars in LMXBs.

To see if the polar spot affected our determination of $q$ and $i$ in 
Section\,\ref{model}, we used our \textsc{XrbSpectrum} model to simulate 
spectra with a polar spot (10 per cent in area) for a given $q$ and $i$; 
we set regions on the star grid near the pole to an effective 
temperature 1000\,K lower than the immaculate, unspotted photosphere. We 
also added instrumental broadening and finally orbital smearing, 
matching that of the observed data. We then produced LSD profiles before 
finally performing a grid search in $q$ and $i$ using our  \textsc{XrbSpectrum} 
model to fit 
the data. We find that we recover the $q$ and $i$ values of our 
simulated data. This is not surprising because a polar spot viewed on a 
system with a low inclination angle will be observable at all orbital 
phases, and it will predominantly only affect the core of the line 
profiles. The determination of $q$ and $i$ using our  
\textsc{XrbSpectrum} model, on the 
other-hand, depend heavily on the {\em asymmetry} of the line profiles, 
and is therefore not greatly affected by features present in the core of 
the lines (see Section\,\ref{dis:lcurve}).

\section{DISCUSSION}
\label{dis}

\subsection{The determination of $q$ and $i$}
\label{dis:lcurve}

\citet{Shahbaz93} modelled the quiescent $H$-band light curve of 
\target\ as due to the Roche-lobe filling secondary star's ellipsoidal 
modulation and obtained $i$ in the range 31$^\circ$--54$^\circ$, assuming no 
contamination from the accretion disk. Using IR spectroscopy 
\citet{Khargharia10} determined the contribution of the secondary star 
to the infrared flux. They then remodelled the $H$-band light curve of 
\citet{Shahbaz93}, correcting for the fractional contribution of the 
donor star to obtain an inclination angle of 35$^\circ$$^{+4^\circ}_{-1^\circ}$
However, as noted by these authors, there are a number of systematic 
effects that may not be accounted for in their uncertainties on $i$. 
Mismatches between template and donor star in temperature, gravity, or 
metallicity would lead to an uncertain determination of the secondary 
star's contribution to the infrared flux. Also it is known that the 
shape and amplitude of the light curve is not consistent throughout 
quiescence, there are various different states characterized by magnitude, colour, 
and aperiodic variability \citep{Cantrell08,Cantrell10}. Determining the 
contribution of the secondary star requires contemporaneous light curves 
and spectroscopy. The presence of starspots introduce systematic errors 
in the determination of $i$ obtained by modelling optical/infrared light 
curves. \citet{Gelino01} observed anomalies in the infrared continuum 
light curve of A\,0620--00, which they modelled as cool star-spots on 
the secondary star. They found that starspots affect the derived 
inclination angle by $\sim$1$^\circ$. In \target\ there is no evidence 
for starspots in the light curves, but this could be due to the fact 
that the light curves are dominated by aperiodic variability 
\citep{Chevalier89, Shahbaz10}. Also, if the dominant spot is located 
over the polar regions, as seen in the Roche tomograms presented in this 
paper, then there will be little rotational modulation due to the spot 
as it is always in view.

\citet{Shahbaz98} assessed the effects of the physical processes that 
determine the shape of the absorption lines in the spectra of Roche-lobe 
filling stars. They found that $i$ primarily affects the degree of 
asymmetry of the line profiles near orbital phases 0.0 to 0.1 and 0.4 to 
0.5 and not the actual widths. The lines become more asymmetric as $i$ 
increases, and $\sim$40$^\circ$ seems to mark the angle where the 
asymmetry of the line profiles are zero. Our determination of $q$ 
and $i$ does not depend on the depth of the absorption lines, and only 
depends on the shape of the phase dependent line profiles.  Starspots 
present near the stellar limb will result in line profiles that appear 
narrower, whereas starspots present near the disc-centre, and/or at high 
latitudes/poles will result in flat-bottomed line profiles which only 
affect the core of the line profiles. For a low inclination system like 
\target, a polar spot is observed at all orbital phases and so will have 
a similar affect on all the phase-resolved line profiles, and given that 
it does not affect the line wings, the polar spot does not affect the 
determination of $q$ and $i$ (see Section\,\ref{model}). Our 
determination of $q$ is accurate to 1.4 per cent because we model the 
exact shape of the rotationally-broadened line profiles in a way that 
does not depend on assumptions about the rotation profile, 
limb-darkening coefficients and the effects of contamination from an 
accretion disk. Previous estimates for $q$ are only accurate to 15 per 
cent \citep{Casares07} mainly because of the uncertainties in the 
limb-darkening coefficient and our determination is a factor of $\sim$10 
more accurate. Our determination of $i$ is accurate to 5$^\circ$, 
similar to that found by previous authors \citep{Khargharia10}, but it 
should be noted that our determination is not affected by aperiodic 
variability or polar starspots.

\subsection{The implications of starspots}

Rotating single stars cooler than about 6500\,K with spectral types 
later than F5 have convective outer layers. They create a dynamo that 
amplifies the internal magnetic fields and brings them to the stellar 
surface. Angular momentum is then removed from the stellar surface by 
the action of the magnetically coupled stellar wind, the mechanism being 
called ``magnetic braking'' \citep{Schatzman62}. The low-mass secondary 
stars in CVs and LMXBs should also have magnetic fields since they also 
have deep convection zones and are rapidly rotating since they are 
tidally locked with their compact companion star. Therefore, magnetic 
braking is thought to be the fundamental mechanism responsible for 
orbital angular momentum loss in CVs and LMXBs, which maintains the mass 
transfer from the low-mass donor to the more massive compact companion 
\citep{Verbunt81, Rappaport83}. Indeed, magnetic braking is crucial in 
the standard explanation for the deficit of CVs with orbital periods 
between 2 and 3\,h (the ``period gap''; \citealt{Robinson81, Rappaport83, Spruit83}).

The secular evolution of LMXBs follows two paths, depending on the 
evolutionary stage of the companion star at the start of the mass 
transfer. \citet{Pylyser88} found that there is a bifurcation period at 
$\sim$12\,h for the initial binary orbital period which separate 
converging and diverging binaries. 
More recent works suggest a bifurcation of 0.5 to 1 d, however its precise
value depends on the treatment of tidal interactions 
and magnetic braking (e.g. \citealt{Sluys05}; \citealt{Ma09}) and
is still a subject of debate.
If the orbital period at the 
beginning of the mass transfer is above the bifurcation period, the 
evolution of the binary begins when the companion evolves off the main 
sequence, the mass transfer is driven by the internal evolution of the 
low-mass (sub-)giant companion star and the system will evolve towards 
large orbital periods.  If the orbital period of the system at the onset 
of the mass transfer is below the bifurcation period, the companion star 
is relatively unevolved, and the only important mechanism driving mass 
transfer is systemic angular momentum loss due to magnetic braking and 
gravitational radiation, which eventually leads to stripped, evolved 
companion stars. For angular momentum loss due to magnetic braking the 
secondary star must have a convective envelope \citep{Rappaport83}, 
which implies that the initial mass of the secondary star has to be less 
than 1.5\,\Msun\ (stars above this mass have radiative envelopes). Our 
evidence for a polar-spot on the secondary star in \target\ (see 
Section\,\ref{spots}) implies that the star must have a convective 
envelope and thus supports the idea that magnetic braking plays an 
important role in the evolution of LMXBs

\section{CONCLUSIONS}

We have determined the  system parameters of the X-ray 
transient \target\ using only phase-resolved high-resolution UVES 
spectroscopy. We first determined the radial-velocity curve of the 
secondary star. We then modeled the shape of the phase-resolved absorption 
line profiles using our X-ray binary model \textsc{XrbSpectrum} that 
computes the exact rotationally-broadened phase-resolved spectrum. This 
approach does not depend on any assumptions regarding the shape of the 
rotation profile, limb-darkening coefficients or laws, or the effects of 
contamination from an accretion disk. We determined the binary mass 
ratio of 0.1755$\pm0.0025$ which is a factor of $\sim$10 more accurate than previous 
estimates. We also constrain the inclination angle to 
32$^\circ$$^{+8^\circ}_{-2^\circ}$. 
Combining these values with the results of the 
radial velocity study gives a neutron star mass of 
1.94$^{+0.37}_{-0.85}$\Msun, consistent with previous estimates.

We present the first Roche tomography reconstruction of the secondary 
star in a LMXB, which reveals the presence of a large, cool polar 
starspot, similar to those seen in Doppler images of rapidly-rotating 
isolated stars. The detection of a starspot supports the idea that 
magnetic braking plays an important role in the evolution of LMXBs.

\section*{ACKNOWLEDGEMENTS}

We would like to thank Tom Marsh for the use of his \textsc{molly} 
software package. We acknowledge the use of the Vienna Atomic Line 
Database (VALD) for obtaining our atomic line lists. This paper makes 
use of the IAC's Supercomputing facility \textsc{condor}. This research 
has been supported by the Spanish Ministry of Economy and 
Competitiveness (MINECO) under the grant (project reference 
AYA2010-18080). CAW and VSD are supported by the STFC.

\footnotesize{

}

\end{document}